\documentclass[10pt,twocolumn]{IEEEtran}

\usepackage{graphicx}
\usepackage{amssymb}
\usepackage{booktabs}
\usepackage{multirow}
\usepackage{array}
\usepackage{tabularx}
\usepackage{balance}
\usepackage{cite}
\usepackage{xcolor}
\usepackage{url}
\usepackage[hidelinks]{hyperref}
\usepackage{tikz}
\usetikzlibrary{arrows.meta,positioning,fit,shapes.geometric}

\definecolor{revisionblue}{RGB}{0,70,180}
\newcommand{\rev}[1]{{\color{black}#1}}

\title{DISCO: Distributed Spectrum Compliance and Orchestration for Scalable IoT Coexistence}

\author{Lyes Saad Saoud and Moussa Ayyash%
\thanks{Moussa Ayyash is with the Department of Computing, Information, and Mathematical Sciences and Technologies, Chicago State University, Chicago, IL 60628, USA (e-mail: mayyash@csu.edu).}}

\begin{document}
\maketitle

\begin{abstract}
\rev{Massive Internet of Things (IoT) deployments increasingly share spectrum with incumbent, licensed, and unlicensed systems under uncertain traffic, fading, mobility, and intermittent coordination. Existing mechanisms, including fixed power limits, listen-before-talk procedures, spectrum access databases, and learning-based resource allocation, address important aspects of coexistence, but they do not provide a common control plane to translate a network-wide interference risk budget into lightweight guidance for many autonomous devices. This article introduces Distributed Spectrum Compliance and Orchestration (DISCO), a hierarchical architecture that separates local spectrum learning from edge-level compliance regulation and slower cloud or non-terrestrial-network context adaptation. DISCO is not presented as a new reinforcement-learning optimizer or as a replacement for statutory spectrum rules. Its contribution is a deployable compliance plane that monitors violation statistics, broadcasts a compact governance signal, and adjusts policy aggressiveness without centralizing every transmission decision. A 30-seed UAV coexistence case study illustrates the efficiency--risk trade-off: the reported mean throughput is 81.0~Mbps, 73\% above fixed-power control, while the mean violation rate is 0.053 compared with 0.126 for uncoordinated learning. Because the 95\% confidence interval, [0.030, 0.076], crosses the nominal target of 0.06, the evidence supports statistical regulation near the target, not guaranteed regulatory compliance. Deployment, complexity, adoption boundaries, and open validation requirements are discussed explicitly.}
\end{abstract}

\begin{figure*}[t]
    \centering
    \includegraphics[width=0.98\textwidth]{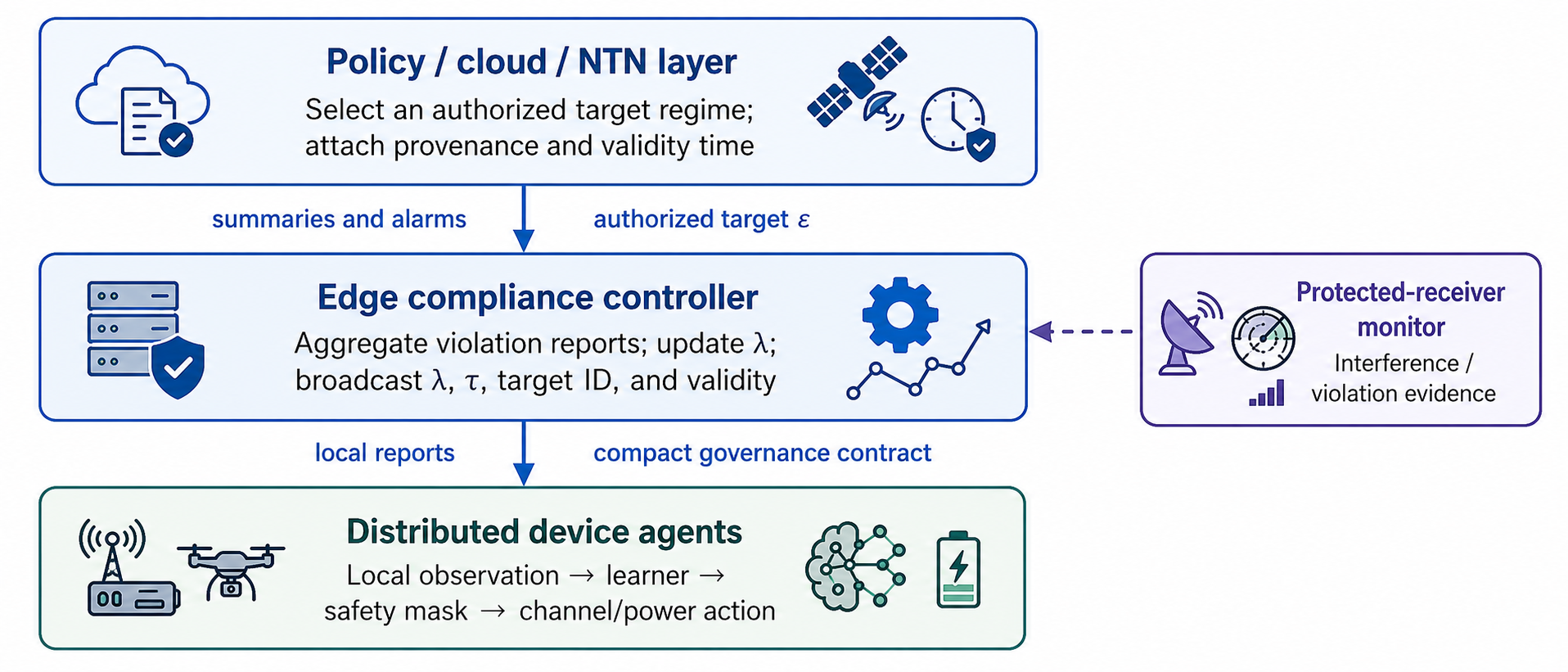}
    \caption{DISCO separates policy ownership, edge compliance regulation, and local spectrum learning. The edge broadcasts compact governance parameters rather than scheduling every transmission. The cloud/NTN layer operates on a slower timescale and cannot override hard regulatory limits.}
    \label{fig:architecture}
\end{figure*}
\section{Introduction}
\rev{Shared-spectrum systems already use several forms of governance. In the U.S. Citizens Broadband Radio Service (CBRS), a Spectrum Access System (SAS) authorizes devices and protects higher-priority users through a three-tier framework \cite{FCCPart96,NISTCBRS}. In 5G New Radio operation in unlicensed spectrum, channel access procedures such as listen-before-talk regulate when a transmitter may occupy a shared channel \cite{ETSI37213}. Open Radio Access Network (O-RAN) architectures add programmable near-real-time and non-real-time control functions, creating practical locations for monitoring and policy applications \cite{ORAN2026}. These mechanisms are valuable, but they solve different problems: SAS grants and coordinates access, listen-before-talk controls short-term contention, and a RAN Intelligent Controller optimizes network behavior. None of them, by itself, specifies how thousands of autonomous IoT learners should collectively remain near an operator-defined interference-risk budget as local conditions and coordination quality change.}

\rev{The scale problem is particularly acute for massive IoT. Large populations transmit asynchronously, observe only local radio conditions, and may use different access technologies. Static power or duty-cycle restrictions are predictable but often conservative. Cognitive-radio and edge-assisted designs improve adaptability \cite{Salameh2018,Hamdan2020,Yang2018}, while distributed multi-agent reinforcement learning (MARL) can learn channel and power decisions without full centralization \cite{Liang2019,Shamsoshoara2019,Jiang2022}. However, unconstrained local optimization can shift interference risk from an individual link to the network as a whole. Constrained Markov decision processes and safe reinforcement-learning methods provide mathematical tools for handling expected costs \cite{Altman1999,Achiam2017}, but an optimization method alone does not define where compliance is measured, who owns the target, how multiple learners receive policy pressure, or how the system degrades when a remote coordination link is unreliable.}

\rev{DISCO addresses this architectural gap. The device learner remains local and algorithm-agnostic. An edge compliance controller converts aggregate violation statistics into a compact governance signal. A slower cloud or non-terrestrial-network (NTN) policy layer selects an operating target from pre-authorized regimes. This separation creates a clear boundary between optimization and governance: devices optimize performance, the edge regulates collective behavior, and the policy plane defines which risk envelope is admissible.}

\rev{The central claim is deliberately narrower than in the rejected version. DISCO does not guarantee legal compliance from simulation, and the projected dual-style update is not itself novel. The novelty is the \emph{distributed compliance plane}: (i) a reusable interface between heterogeneous local learners and a network-wide risk budget; (ii) a separation of fast device actions, medium-timescale edge regulation, and slow policy adaptation; and (iii) an operational pathway that can coexist with hard regulatory mechanisms rather than replacing them.}

\rev{This article makes three contributions. First, it positions DISCO against concrete spectrum-control mechanisms and clarifies what is new and what is inherited from prior work. Second, it specifies the device observation/action/reward interface, the safety mask, the edge feedback loop, and the role of the NTN context indicator at a level suitable for system implementation. Third, it reinterprets the existing case study conservatively, adds deployment and complexity analysis, and states the evidence boundary without introducing new or fabricated results.}

\section{Where DISCO Fits in the Spectrum-Control Stack}

\begin{table*}[t]
\centering
\rev{\caption{Positioning of DISCO relative to representative spectrum-control mechanisms. The comparison is architectural; it does not imply that the mechanisms are mutually exclusive.\label{tab:positioning}}}
\rev{\small
\begin{tabularx}{0.98\textwidth}{p{2.35cm}p{2.35cm}p{2.25cm}X X}
\toprule
\textbf{Mechanism} & \textbf{Primary decision} & \textbf{Typical timescale} & \textbf{Strength} & \textbf{Gap addressed by DISCO} \\
\midrule
Fixed power/duty-cycle rule & Device operating limit & Configuration time & Predictable and simple to audit & Conservative under changing density and channel conditions \\
CBRS SAS & Authorization, channel and power grants & Database/control transactions & Protects priority tiers through an established regulatory framework & Does not continuously shape the joint behavior of heterogeneous local learners \\
NR-U shared-channel access & Whether/when a node may transmit & Slot/channel-access time & Technology-defined contention discipline & Does not expose a network-wide probabilistic risk budget to autonomous policies \\
Constrained or safe RL & Policy optimization subject to a cost & Training and decision time & Directly couples reward and constraints & Usually binds the constraint to one training formulation rather than a reusable governance plane \\
O-RAN RIC & Programmable monitoring and RAN control & Near-real-time and non-real-time & Practical control placement and open interfaces & Requires an application-specific compliance contract and aggregation rule \\
\textbf{DISCO} & Collective policy pressure and admissible risk regime & Edge window plus slower policy update & Separates local learning, compliance monitoring, and policy ownership & Must still be integrated with hard rules, certified measurements, and audited deployment procedures \\
\bottomrule
\end{tabularx}}
\end{table*}

\rev{DISCO should therefore be viewed as an overlay that complements existing controls. For example, a CBRS device would still require valid SAS authorization, and an NR-U transmitter would still follow the applicable channel-access procedure. DISCO acts inside those hard boundaries, regulating how aggressively autonomous devices exploit their permitted choices. Likewise, an O-RAN deployment could host the edge controller as an xApp or equivalent edge service and host slower target selection in the non-real-time control plane.}

\section{Architecture and Operational Contract}

\subsection{Device-Level Learning and Safety Filtering}

\rev{Each IoT or UAV device maintains a local spectrum-access policy. The architecture does not require a particular optimizer: a value-based learner, actor--critic method, contextual bandit, or rule-based adaptive policy can use the same interface. In the reported case study, the relevant local decision is a candidate channel/power transmission action. The observation contains locally available context such as sensed channel occupancy, received signal or interference measurements, position/context information, and the most recently received governance parameters.}

\rev{The local objective is high communication utility, represented in the case study by link throughput. The edge governance signal adds a cost to actions associated with interference violations. This is conceptually related to a Lagrangian penalty, but DISCO treats the penalty as a network service rather than embedding one fixed dual variable inside one learner. Consequently, devices can use heterogeneous learning algorithms while responding to the same compliance pressure.}

\rev{Before execution, a safety mask removes candidate actions whose estimated interference risk exceeds a configured threshold $\tau$. The mask is a preventive screen, not a proof of zero interference. A practical implementation must define a conservative fallback, for example, minimum permitted power, a protected channel, or no transmission, when no candidate survives. Hard technology and regulatory constraints always take precedence over the learned policy and the DISCO signal.}

\subsection{Edge Compliance Controller}

\rev{The edge layer is the architectural core. Devices and, when available, protected-receiver monitors send compact statistics rather than raw high-rate observations. Over a sliding window, the edge estimates the empirical violation rate $\bar v$. A violation is the positive output of the simulator's configured interference-threshold test; the reported violation rate is the empirical fraction of evaluated decisions marked as violations. The edge compares $\bar v$ with the active target $\varepsilon$. If the rate is above target, it increases the nonnegative governance variable $\lambda$ in proportion to the gap; if the rate is below target, it relaxes $\lambda$ toward zero. The update is projected at zero so that the signal cannot become a reward for risky behavior.}

\rev{Two implementation parameters determine responsiveness: the observation-window length and the update step size. A longer window reduces noise but delays reaction; a larger step reacts faster but can oscillate. The original simulation record supplied with this manuscript does not contain the numerical values of these two parameters. They are therefore not invented here, and no reproducibility claim is made for them. A release intended to support exact replication must publish these values, the aggregation rule, the threshold test, and the simulator code.}

\rev{The downlink control payload is small: the edge broadcasts $\lambda$, the safety threshold $\tau$, the active target identifier, and a validity time. The architecture avoids per-device centralized scheduling. Its monitoring work grows approximately linearly with the number of device reports in each window, while the governance update itself is constant-size. These are structural complexity observations, not measured runtime or signaling-overhead results.}

\subsection{Cloud/NTN Context Adaptation}

\rev{The rejected manuscript used the phrase ``satellite visibility,'' which can mean geometry, elevation angle, coverage probability, or link quality. In this article, the term is replaced by \emph{NTN control-link quality/availability}. It denotes a coarse context state indicating whether the remote coordination path is nominal or degraded. The present case study does not simulate orbital geometry, satellite handover, propagation delay, or a standard-specific NTN link budget. ``NTN shadowing'' is therefore a scenario label for degraded observability/coordination, not a claim of a complete satellite-channel model.}

\rev{The cloud/NTN layer operates more slowly than the edge. It selects among pre-authorized targets, here $\varepsilon=0.06$ in the nominal regime and $\varepsilon=0.10$ in the degraded regime. In a real deployment, this layer must never relax a statutory interference limit autonomously. A higher numerical target is admissible only when it represents an operator-defined internal metric within an already legal envelope, or when the metric changes because measurement confidence has changed. Policy provenance, authorization, and expiration must be logged.}

\subsection{What Is New---and What Is Not}

\rev{Projected dual updates, expectation constraints, safety masks, edge computing, and MARL are established ideas \cite{Altman1999,Achiam2017,Liang2019}. DISCO's contribution is their system-level organization into a three-timescale compliance service. First, the edge owns a network-wide violation statistic rather than each learner maintaining an isolated constraint. Second, the interface is learner-agnostic: heterogeneous devices receive the same compact governance contract. Third, the policy plane changes the target regime without entering the fast action loop. Fourth, the architecture explicitly supports audit data, validity time, conservative fallback, and coexistence with hard spectrum rules. These properties distinguish DISCO from simply applying constrained MARL to another resource-allocation problem.}

\section{Case Study and Evidence Boundary}

\subsection{Scenario and Compared Configurations}

\rev{The existing case study uses a $1000\times1000$~m$^2$ urban micro-cell, eight primary users, six channels, and secondary UAV agents. Each run contains 200 training episodes with 50 decisions per episode, yielding 10,000 decision steps. Episodes 100--199 apply the degraded NTN-context condition; thus the transition occurs at decision step 5,000. The nominal setting has more primary users than channels, creating contention without the 50\% increase used in the stress setting, where the primary-user count rises from eight to twelve. These choices are suitable for an architecture stress illustration but are not claimed to represent one specific commercial deployment.}

\rev{Three configurations are reported. The deterministic baseline uses fixed 10~dBm power and serves as a conservative reference. The no-edge ablation uses the same local learning mechanism without compliance feedback, isolating the effect of the governance plane. DISCO combines local learning, the safety mask, edge feedback, and the two context targets. Stronger algorithmic baselines such as constrained policy optimization, Lagrangian MARL, interference-aware heuristics, or an O-RAN xApp are important future comparisons, but numerical results for them do not exist in the supplied experiment and are not fabricated. Table~\ref{tab:positioning} provides only a conceptual comparison.}

\rev{The reported throughput is the simulator's aggregate secondary-network throughput, and the violation statistic is produced by its fixed interference-threshold test. The source package does not include the simulator implementation, channel equation, mobility trace, absolute interference threshold, or detailed signaling model. Consequently, the numerical results support comparison among the three configurations under one common simulator; they do not establish 3GPP conformance, FCC compliance, absolute field performance, or independent reproducibility. This boundary is a weakness, but stating it is scientifically preferable to inserting a channel model after the experiments were completed.}

\rev{Thirty independent random seeds are reported. The 95\% confidence interval around a sample mean is computed using Student's $t$ distribution: the sample mean plus or minus the $0.975$ quantile with 29 degrees of freedom multiplied by the sample standard deviation divided by $\sqrt{30}$. This interval quantifies uncertainty across the simulated seeds; it is not a regulatory confidence guarantee.}

\begin{table}[t]
\centering
\rev{\caption{Reported case-study results. No new numerical experiment has been added.\label{tab:validation}}}
\rev{\small
\resizebox{\columnwidth}{!}{%
\begin{tabular}{lccc}
\toprule
\textbf{Configuration} & \textbf{Violation rate} & \textbf{Throughput} & \textbf{Enforcement} \\
 & & \textbf{(Mbps)} & \textbf{density} \\
\midrule
Fixed 10~dBm & 0.022 & 46.8 & -- \\
No-edge RL & $0.126\pm0.064$ & 105.2 & -- \\
DISCO, nominal & 0.053 & 81.0 & 0.545 \\
DISCO, stress & 0.082 & 71.4 & 0.705 \\
\bottomrule
\end{tabular}}}
\end{table}

\subsection{Efficiency--Risk Trade-Off}

\rev{The fixed-power reference records the lowest mean violation rate, 0.022, but also the lowest throughput, 46.8~Mbps. Removing edge governance raises throughput to 105.2~Mbps while increasing the violation rate to 0.126. Under nominal conditions, DISCO records 81.0~Mbps and a violation rate of 0.053. The throughput gain relative to fixed power is therefore approximately 73\%, and the violation rate is approximately 2.4 times lower than the no-edge ablation. These comparisons support the intended architectural effect: the edge signal sacrifices part of the unconstrained learner's throughput to reduce collective interference exposure.}

\rev{The correct interpretation of the confidence interval is important. For the nominal DISCO violation rate, the mean is 0.053 and the reported 95\% interval is [0.030, 0.076]. Because the upper bound exceeds the target of 0.06, the experiment does not demonstrate target satisfaction at 95\% confidence and must not be described as guaranteed or certified compliance. It indicates regulation around the target across 30 seeds. Establishing a stronger claim would require more runs, a pre-specified statistical test, tail-risk analysis, and a complete measurement definition.}

\begin{figure}[t]
\centering
\includegraphics[width=0.96\columnwidth]{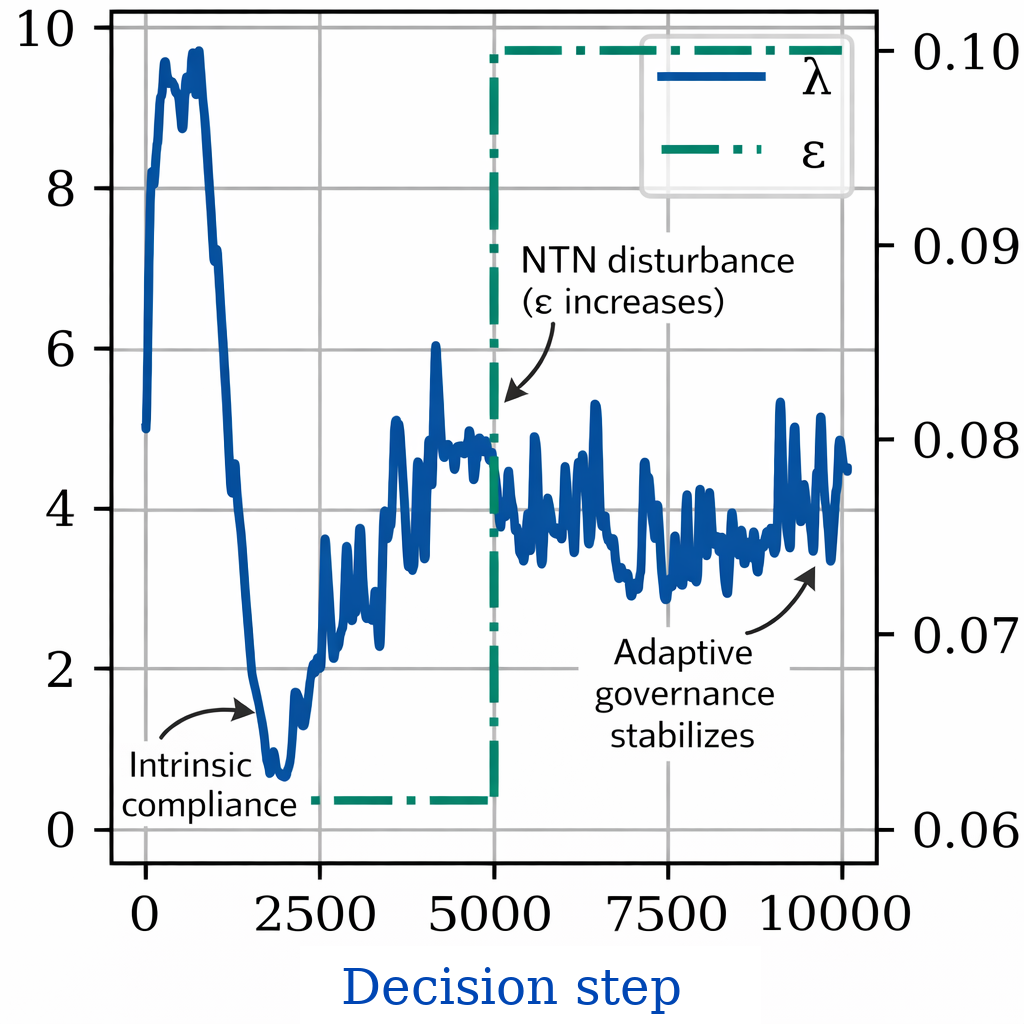}
\rev{\caption{Governance signal and active target across 10,000 decision steps. The target changes at step 5,000, corresponding to the start of the degraded-context episodes. The trajectory illustrates feedback adaptation; it is not evidence of instantaneous constraint satisfaction.\label{fig:governance_dynamics}}}
\end{figure}

\rev{Figure~\ref{fig:governance_dynamics} shows that the governance signal can relax when the observed statistic is below the active target and rise when pressure increases. The figure's original horizontal label, ``Episode,'' was corrected to ``Decision step'' because the simulation contains 200 episodes of 50 steps, not 10,000 episodes. This correction aligns the visualization with the stated experiment.}
\begin{figure*}[!ht]
    \centering
    \includegraphics[width=0.9\textwidth]{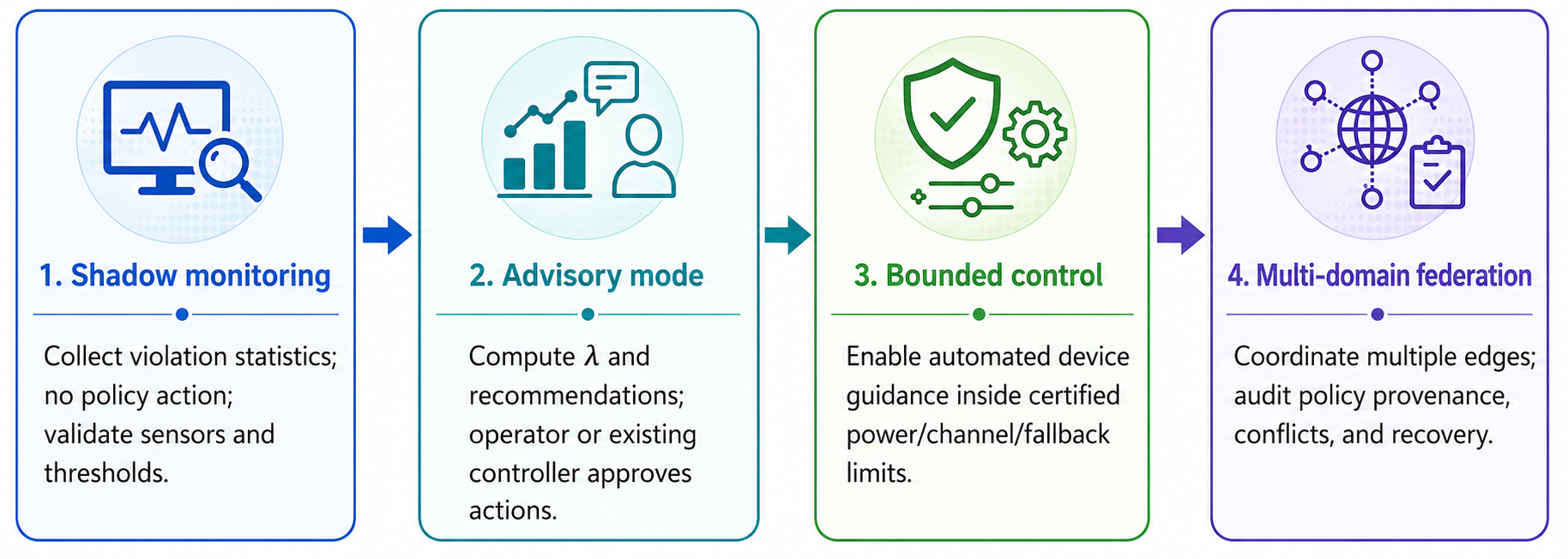}
    \caption{Recommended adoption sequence. DISCO should progress from observation to advisory operation before receiving bounded actuation authority; federation is a later step, not a starting assumption.}
    \label{fig:deployment}
\end{figure*}
\subsection{Stress and Sensitivity Results}

\rev{Increasing the primary-user count from eight to twelve raises the reported violation rate from 0.053 to 0.082 and the enforcement density from 0.545 to 0.705, while throughput decreases from 81.0 to 71.4~Mbps. The result is consistent with a controller that applies more pressure under denser coexistence. However, no confidence interval is supplied for the stress mean, so the value should not be used to claim statistically verified satisfaction of the relaxed target.}

\rev{The available target sweep uses $\varepsilon\in\{0.04,0.06,0.08\}$. Enforcement density decreases monotonically from 0.885 to 0.664 and 0.599 as the target is relaxed. This is evidence that the governance mechanism responds in the expected direction. The reported correlation between target and realized violation rate is only $r=0.12$ for the three tested settings; with only three target points, this value is descriptive and should not be treated as a calibration result. Sensitivity to window length, update step size, device density beyond the one stress point, and severity/duration of the degraded NTN context remains untested in the supplied data.}

\rev{The supplied severe-shadowing trace also exposes a key limitation of feedback governance: a violation increase may precede a rise in $\lambda$ because measurements are aggregated over a finite window. This lag is why DISCO cannot replace hard masks, maximum-power limits, emergency shutdown, or incumbent-protection procedures. It is also why future evaluation should report peak violation duration and upper-tail metrics in addition to a mean rate.}

\section{Complexity, Deployment, and Adoption}

\subsection{Training and Runtime Complexity}

\rev{DISCO does not require centralized joint training. Each device pays the computational cost of its chosen local policy, which is unchanged between the no-edge ablation and the full architecture except for receiving the governance parameters and applying a risk cost/mask. The edge receives one compact report per participating device per reporting interval; aggregation therefore grows linearly with the number of reports. After aggregation, updating one scalar governance signal and one target state is constant-size work. Broadcasting the same small contract to all devices can use multicast/broadcast where available or unicast replication otherwise.}

\rev{These statements describe scaling order only. The current experiments do not measure processor time, memory, message size, backhaul load, or control-loop latency. A deployability study must provide those measurements and test stale, delayed, duplicated, and missing reports. It should also compare centralized aggregation with hierarchical multi-edge aggregation when a single edge domain becomes too large.}

\subsection{Deployment Mapping}

\rev{A realistic mapping has four roles. First, device software or firmware hosts the local learner and safety mask. Second, an edge gateway, private network controller, or near-real-time O-RAN application hosts interference aggregation and the governance update. Third, a non-real-time controller or operator cloud manages authorized targets, model versions, policy provenance, and audit records. Fourth, existing enforcement systems, such as SAS authorization, channel access rules, certified power limits, and incumbent protection functions, remain outside DISCO and retain priority.}

\rev{Figure~\ref{fig:deployment} proposes a staged rollout. Shadow monitoring is essential because a simulator-defined violation indicator may not transfer directly to field measurements. Advisory operation then tests whether the governance signal is understandable and stable without granting actuation authority. Bounded control is appropriate only after the fallback, validity horizon, maximum actuation, and audit path are certified. Multi-edge federation should follow only after conflict resolution and policy ownership are defined.}

\subsection{Regulatory and Operational Boundaries}

\rev{The phrase ``probabilistic compliance'' can be misleading if it suggests that a legal interference limit may be violated with some accepted probability. DISCO instead regulates an operator-defined statistical metric \emph{inside} hard legal and technology-specific limits. A regulator, SAS administrator, or standards-based access mechanism defines the admissible envelope. DISCO helps allocate performance within that envelope and provides an audit trail showing how collective policy pressure changed.}

\rev{This distinction also changes the deployment claim. DISCO is not ready for autonomous real-world adoption based on the current simulation. It is an architecture and an initial case study. Before field use, the violation metric must be tied to calibrated measurements at protected receivers; the target must have a documented legal/operational interpretation; the safety mask and fallback must be tested under communication loss; and the entire control chain must be assessed for cybersecurity, false reports, and policy manipulation.}

\section{Open Research Issues}

\rev{\textbf{Explainability and auditability:} Operators need a trace linking an observed interference event to contributing devices, the active target, the governance signal, and the action filters applied. A scalar $\lambda$ is compact but not self-explanatory. Attribution and counterfactual diagnostics are needed.}

\rev{\textbf{Scalability and dependence:} Reports from nearby devices are correlated, so treating all samples as independent can make a confidence interval overoptimistic. Multi-edge aggregation must handle overlapping interference domains without double counting or conflicting penalties.}

\rev{\textbf{Tail risk and reaction time:} Mean violation rate is insufficient for incumbent protection. Future work should report maximum excursion, duration above threshold, conditional tail expectation, and recovery time, and should evaluate risk-sensitive or tightened targets when feedback delay is high.}

\rev{\textbf{Security and trust:} A malicious device could under-report interference, replay stale governance parameters, or manipulate the edge estimate. Signed reports, freshness checks, robust aggregation, and fail-closed behavior are necessary for a trustworthy compliance plane.}

\rev{\textbf{Baseline and reproducibility gap:} The next experimental version should release the channel/mobility model, reward, action discretization, threshold, safety-mask rule, window length, step size, and seeds. It should compare fixed power, an interference-aware heuristic, constrained RL, Lagrangian MARL, and an edge/O-RAN controller under equal information and compute budgets. Hardware-in-the-loop or software-defined-radio evaluation is needed before making adoption claims.}

\rev{\textbf{Policy semantics:} A target such as 0.06 is meaningless without a measurement unit, population, aggregation window, protected entity, and consequence. A deployment contract should define all five and forbid remote relaxation beyond an authorized envelope.}

\section{Conclusion}

\rev{DISCO introduces a distributed compliance plane for IoT spectrum coexistence. Its architectural contribution is not a new RL algorithm or a new dual update; it is the separation of heterogeneous local learning, edge-owned statistical regulation, and slower policy-context adaptation. This separation makes the control responsibility explicit and offers a practical mapping to edge gateways or programmable RAN controllers while preserving the priority of existing spectrum rules.}

\rev{The supplied 30-seed case study shows a meaningful efficiency--risk trade-off: 81.0~Mbps under nominal DISCO operation, approximately 73\% above fixed 10~dBm control, and a violation rate approximately 2.4 times lower than uncoordinated learning. The evidence must be interpreted carefully. Because the nominal 95\% confidence interval crosses the target, the results demonstrate regulation near a target rather than guaranteed compliance. The missing simulator, channel, threshold, and control-parameter details also prevent independent reproduction and field-level claims.}

\rev{The next step is therefore not stronger wording but stronger evidence: complete simulator disclosure, competitive constrained-learning baselines, tail-risk metrics, measured overhead, and staged testbed validation. With those additions, DISCO could mature from a promising architecture into an auditable component of self-governing shared-spectrum networks.}

\balance

\end{document}